\documentclass[twocolumn,showpacs,nofootinbib]{revtex4-1}
\usepackage{epsfig,amssymb,amsmath,latexsym}
\usepackage{pifont}
\usepackage{hyperref}

\usepackage{color}
\definecolor{nicered}{rgb}{0.7,0.1,0.1}
\definecolor{nicegreen}{rgb}{0.1,0.5,0.1}

\newcommand{\be}{\begin{equation}}
\newcommand{\ee}{\end{equation}}
\newcommand{\bea}{\begin{eqnarray}}
\newcommand{\eea}{\end{eqnarray}}

\newcommand{\no}{\noindent}
\newcommand{\nb}{\nonumber}

\newcommand{\de}{\partial}

\newcommand\Q{{\cal Q}}
\renewcommand\S{{\cal S}}
\newcommand\T{{\cal T}}

\renewcommand\l{\lambda}

\renewcommand\a{\alpha}
\newcommand\D{\text{DoF}}
\renewcommand\b{\beta}
\renewcommand\k{\kappa}

\newcommand\g{\gamma}

\newcommand\V{{\ensuremath{\cal V}}}

\renewcommand\l{\ensuremath{\lambda}}

\newcommand\ba{\begin{array}}
\newcommand\ea{\end{array}}

\newcommand{\plm}{M_{\text{pl}}^2}

\newcommand\SEC[1]{\medskip\noindent{\sl\bfseries #1}}


\begin{document}

\title{Degrees of Freedom in Massive Gravity}

\author{D. Comelli$^a$, M. Crisostomi$^{b,c}$, F. Nesti$^b$ and   L. Pilo$^{b,c}$}
\affiliation{
  $^a$INFN - Sezione di Ferrara,  I-35131 Ferrara, Italy\\
  $^b$Dipartimento di Fisica, Universit\`a di L'Aquila, I-67010 L'Aquila\\
  $^c$INFN, Laboratori Nazionali del Gran Sasso, I-67010 Assergi, Italy
}
\date{\small \today}

\begin{abstract}
  \no We study in a systematic way a generic non derivative (massive) deformation of general
  relativity using the Hamiltonian formalism.  The number of propagating degrees of freedom is
  analyzed in a non-perturbative and background independent way.  We show that the condition of
  having only five propagating degrees of freedom can be cast in a set of differential equations for
  the deforming potential.  Though the conditions are rather restrictive, many solutions can be
  found.
\end{abstract}


\maketitle

\no Whether general relativity (GR) is an isolated theory is an interesting question both from the
theoretical and phenomenological side.  In particular, attempts to modify GR at large distances
recently have received a renewed interest.  In this paper we focus on massive deformation of GR.  A
possible  definition of Lorentz-invariant (LI) massive gravity is any deformation of GR such that
the metric perturbation $h_{\mu\nu}$ around flat Minkowski space $\eta_{\mu\nu}$ behaves as a
massive spin 2 particle with 5 polarizations (DoF) in contrast with GR where only 2 DoF propagate.

Since the work of Fierz and Pauli (FP)~\cite{Fierz:1939ix} it was realized that for a generic
massive deformation, even at linearized level 6 DoF are present. Though a tuning allows to get rid
of the sixth mode, it reappears at the non-linear level or around non flat backgrounds~\cite{BD}. In
addition the sixth mode is typically a ghost.  Recently, it was shown that there exists a LI non
linear completion of the FP theory that is free of ghost up to the fourth
order~\cite{Gabadadze:2011} avoiding the presence of the sixth mode. Then proof of the propagation
of only five DoF and the absence of instabilities was extended to the non perturbative level
in~\cite{GF,DGT}. Cosmology of such a model was studied in~\cite{cosm}.  On the other hand, one can
give up Lorentz invariance in the gravitational sector, and a class of theories of massive gravity
that avoid the vDVZ discontinuity~\cite{DIS} is available at quadratic
level~\cite{Rubakov,dub,PRLus,gaba}.

The main goal of this paper is to study the number of propagating modes in generic massive
deformations of GR, non perturbatively and in a background independent
way. We stress that our analysis applies to both the Lorentz invariant
and Lorentz breaking case. For a different approach that focus only on
the Lorentz invariant potential in \cite{Gabadadze:2011} see~\cite{Kluson}.  

\medskip

The natural tool for such a task is the Hamiltonian formulation in the Arnowitt-Deser-Misner
(ADM)~\cite{ADM} formalism. The 10 DoF of the metric are split in the lapse $N$, the shift $N^i$ and
the spatial 3-metric $\g_{ij}$ ($i,j$=1,2,3)
\be
g_{\mu \nu} = \begin{pmatrix} - N^2 + N^i N^j \g_{ij} & \g_{ij}N^j \\
\g_{ij}N^j  & \g_{ij} \end{pmatrix}.
\ee
Any massive deformation of gravity can be obtained by adding to the Einstein-Hilbert action an
arbitrary non-derivative function $ V (N, N^i, \g_{ij})$ of these variables. Thus the
Hamiltonian reads
\be
H= \plm \int d^3x \left[N^A \,  {\cal H}_A 
+ \, m^2 \,  N \, \sqrt{\g} \, V  \right] \, ,
\label{act}
\ee
where we introduced the notation $N^A=N_A=(N, \,N^i)$ and ${\cal H}_A = ({\cal H}, \,{\cal H}_i)$,
with $A=0,1,2,3$.  The derivative $\partial F/\partial N^A$ of a function $F$ with respect to $N^A$
will be denoted by $F_{,A}$, and we define $\V= m^2 \, N \, \g^{1/2}
\, V $, with $\gamma = \det(\gamma_{ij})$.   As matter of fact,
${\cal H}_A$ are functions of the spatial metric $\g_{ij}$ and its conjugate momenta $\Pi^{ij}$
only.

When $m=0$, the action (\ref{act}) reduces to the familiar ADM GR Hamiltonian derived from the
Einstein-Hilbert action. In general, this action can describe Lorentz invariant (LI) or Lorentz
breaking (LB) models~\cite{Rubakov,dub}. 

Though $\V$ in (\ref{act}) breaks diffeomorphisms, by adding a suitable set of St\"uckelberg scalar
fields gauge invariance can be restored~\cite{HGS} and (\ref{act}) can be considered as a generally
covariant theory written in a unitary gauge.

Peculiar of GR and of its deformed version (\ref{act}), is that $N^A$ are non dynamical and that
their conjugate momenta $\Pi_A$ are vanishing on the physical phase space, which are called primary
constraints. The canonical analysis of constraints is thus appropriate to assess non-perturbatively
the number of degrees of freedom DoF.  

\SEC{Canonical analysis.} The total Hamiltonian is
\be
H_T = H + \int d^3x \,  \lambda^A \Pi_A \, ,
\ee
where a set of Lagrange multipliers $\lambda^A$ are introduced to enforce the primary constraints
$\Pi_A \approx 0$.\footnote{The customary notation $\Pi_A \approx 0$ means that $\Pi_A$ are only
  weakly zero, e.g. zero on the constraints surface only.}  The time evolution of any function of
$\g_{ij}$, $N^A$ and their momenta is given by the Poisson bracket with $H_T$:
\be
\begin{split}
\frac{d f(t,x)}{dt} =&\, \{ f(t,x), H_T(t) \} = \{ f(t,x), H(t)\}+{}\\
& {}+ \int d^3 y \, \lambda^A(t,y)
\{F(t,x), \Pi_A(t,y) \} \, .
\end{split}
\ee
The Poisson bracket between two functionals $F$ and $G$ of canonical variables $q_a,p^a$ is
defined as
$$
\{F, \,G \} 
=\! \int \!d^3 y \left [ \frac{\delta F}{\delta q_a(t,y)} \frac{\delta
  G}{\delta p^a(t,y)} -  \frac{\delta F}{\delta p_a(t,y)} \frac{\delta
  G}{\delta q^a(t,y)} \right] .
$$
In the following the dependence on space and time of the various fields will be understood.

Even before studying the equations of motion of the dynamical variables $h_{ij}$, it is crucial to
enforce that the four primary constraints are conserved in time. This leads to secondary
constraints $\S_A$:
\be
 \S_A\equiv\{\Pi_A, H_T \} =- ({\cal H}_A + \V_{,A}) \approx 0 \, .
\label{second}
\ee 
Basically, these are the equations of motion of $N^A$. They do not determine any of the Lagrange
multipliers and are not automatically satisfied. Therefore, $\S_A\approx 0$ reduce the dimension of
the physical phase space by 4. Also $\S_A$ must be conserved, leading to the new conditions 
\bea
\label{ter}
&&\T_A \equiv \{ \S_A, \, H_T \} = \{ \S_A, \, H \}
- \V_{AB} \, \lambda^B \approx 0 \, ,\\[1ex]
&& \qquad \V_{AB}\equiv \V_{,AB}=\frac{\de^2 \V}{\de N^A \de N^B}\, . \nb
\eea
Now, if the Hessian $|\V_{AB}|$ is non-degenerate, it can be inverted and the four eqs.~(\ref{ter})\
can be solved for all the $\lambda^A$.
At this point all constraints are consistent with time evolution and the procedure stops.

Summarizing, in case $\det|\V_{AB}| \neq 0$ we have $4\, (\Pi_A)+ 4\, (\S_A)=8$ constraints, for a
total of $(20-8)/2=6$ DoF.  Technically, these 8 constraints are Second Class, being Rank$|\{
\Pi_A,\,\S_B\}|=$ Rank$|\V_{AB}|=4$. As a result, the whole  gauge
invariances of GR is  broken.

In the LI case, expanding around a Minkowski background, Lorentz invariance tells us that the
propagating DoF must be grouped  in a massive spin two (5 DoF)
representation plus a scalar (1 DoF).  This
is the Boulware-Deser result, valid for a generic potential.  The extra scalar, the so called
Boulware-Deser sixth mode~\cite{BD}, turns out to be a ghost rendering a generic nonlinear
completion unviable.

The first important result is thus that a necessary condition to have less than six propagating DoF,
is that $r \stackrel{\text{def}}{\equiv}\text{Rank}|\V_{AB}|<4$.

\SEC{The case  $r=3$.} In this important case the matrix ${\cal V}_{AB}$ has a null eigenvector
$\chi^A$ and three other eigenvectors $E_n^A$ $(n=1,2,3)$ with eigenvalues $\k_n$:
\be
\V_{AB} \, \chi^B = 0\,,\qquad\V_{AB} \, E_n^B = \k_n \, E_n^A\, .
\label{null}
\ee
For instance if $\det({\cal V}_{ij}) \neq 0$, then $\chi^A=(1, - {\cal V}^{-1}_{ij} \, {\cal V}_{0
  j})$.  The Lagrange multipliers can also be split along $\chi^A$ and its orthogonal complement
\be
\lambda^A = z \, \chi^A + \sum_{n=1}^{3}\, d_n \,E^A_n \stackrel{\text{def}}{\equiv}  z \, \chi^A +{\bar
  \lambda}^A\, .
\label{decomp}
\ee 
The Hessian is now non-degenerate only in the subspace orthogonal to $\chi^A$, and from (\ref{ter})\
one can determine only three out of four Lagrange multipliers, $d_n = E^A_n\{ \S_A, \,H \}/\k_n$.
The projection along $\chi^A$ leaves $z$ undetermined and~(\ref{ter}) gives a new non-trivial
(tertiary) constraint
\be
\T \equiv \chi^A \, \{\S_A, H \} \approx 0 \, . 
\ee  
Here and in the following we suppose that the constraints found are non trivial.  At this point,
either the conservation of $\T $ allows to determine the remaining multiplier $z$ or it may generate
a further constraint. One has
\bea
\label{quat}
&\Q = \{ \T, H_T \} =  \{ \T, H \} + \{ \T,
\lambda^A \cdot \Pi_A \} \, , \\[1ex]
\nb
&\lambda^A \cdot \Pi_A \equiv \int d^3y \, \lambda^A(y) \Pi_A(y) \, .
\eea
Clearly, the first piece of (\ref{quat}) does not contain $z$. Using (\ref{null})\ and
(\ref{decomp})\ after a tedious but straightforward computation we find for the second piece
\bea\nonumber
  \{ \T,
\lambda^A \cdot \Pi_A \} &\approx&  \chi^B \{ {\bar
  \lambda}^A \cdot \S_A, \S_B  \} -  \chi^B \{  {\bar \lambda}^A {\cal
  V}_{BA} , H \} \\
&&
+\frac{\de \chi^B}{\de N^A}\, {\bar \lambda}^A \,{\bar \lambda}^D\, \V_{BD} - \Theta \cdot  z \,,
\eea
\vspace*{-4ex}
\be
\Theta(x,y) =\chi^A(x)\,  \{S_A(x),S_B(y) \} \, \chi^B(y) \, .
\ee
It turns out that $\Theta(x,y) = A^i(x,y) \de_i \delta^{(3)}(x-y) $ with $A(x,y)=A(y,x)$; it is
worth to point out that in a finite dimensional context $A^i$ would be absent. Thus,
\be
\Theta \cdot z = \int \!d^3 y \,\, \Theta(x,y) z(y) 
=  - \frac{1}{2z(x)}\de_i \left[ z(x)^2 A^i(x,x) \right].
\ee
As a result, $Q$ is free from $z$ if $A^i(x,x)=0$, which consists in the following new condition
\be
{\chi^0}^2 \,  \,\tilde  {\cal V}_i  + 2  \, \chi^A  \chi^j
\,\frac{\de \tilde  {\cal V}_A}{\de \gamma^{ij}} \,
 =0  \, , \qquad \tilde{\cal V}\equiv \gamma^{-1/2}    {\cal V} \, .
\label{add}
\ee
If this condition is satisfied, $\Q$ in (\ref{quat})\ is a quaternary constraint, which together
with $\T$ confirms the elimination of one DoF.  If condition (\ref{add}) is not satisfied,
(\ref{quat}) is not a constraint but can be used to determine the last Lagrange multiplier $z$. This
leaves us with an odd number of second class constraints and an odd dimensional physical phase
space, in particular $5+1/2$ DoF.  A similar situation is encountered in other deformations of GR,
see for instance~\cite{Henneaux:2010vx}.  It not clear if such a peculiar occurrence is physically
acceptable or not. For sure, condition (\ref{add}) is rather
restrictive.

When (\ref{add}) is satisfied by ${\cal V}$, the time evolution of the new constraint $\Q$ is non
trivial and can finally determine the multiplier
\be
z = -\left(\chi^A \,\Q_{,A} \right)^{-1} \left[ \,\bar \lambda^B \,\Q_{,B} +\{\Q, H \} \right]\,.
\ee
In conclusion, for $r=3$, there are $4\, (\Pi_A)+ 4 \, (\S_A) + 1\, (\T)+1 \, (\Q)=10$ constraints,
giving $ (20-10)/2 =5$ DoF, that is a good candidate for a theory of a massive spin-2.  The
required conditions are that ({\ref{add})\ holds and
\be
{\rm Rank}|\V_{AB}|=3\;\;\;\  {\rm and}\;\;\; \ \chi^A \, \Q_{,A} \neq 0\to \D=5.
\ee
To our knowledge, the observation that a singular Hessian is a necessary condition to avoid the
Boulware-Deser argument which leads to six propagating DoF was made for the first time in general
terms in~\cite{DGT}. 

We stress that the condition (\ref{add}) does not appear at the linearized level around flat space;
indeed, the dependence on $z$ starts at the quadratic level thus it manifests either at the
non-linear level or in a non trivial background.  As a result, a random extension of the non-linear
level of the quadratic action of \cite{Rubakov} will not satisfy (\ref{add}).

\SEC{General case.}  Let us briefly discuss the generic case with the Hessian being of any
Rank$|\V_{AB}|=r$ between zero and four.  Now $\V_{AB}$ has $4-r$ null eigenvectors $\chi^A_\alpha$
\be
\V_{AB}\,\chi^B_\alpha=0\,,\qquad \alpha=1,\dots,4-r\,,
\ee
and as before, we can split the Lagrange multipliers as
\be
\lambda^A=\sum_{\alpha=1}^{4-r}\,z_\alpha\,
\chi_{\alpha}^A+\sum_{n=1}^{r}\, d_n \, E^A_n \, .
\ee
The projections of (\ref{ter})\ along $E^A_n$ determine the $r$ Lagrange multipliers $d_n=
E^A_n\{\S_A,H\}/\k_n$, while the projections along the $\chi_\alpha^A$ give $4-r$ tertiary
constraints
\be
\T_\alpha \equiv \chi_\alpha^A\,\{\S_A,H\}\approx 0\,.
\ee
The conservation of these contraints leads to
\be
\Q_\alpha= \{ \T_\alpha,H\} 
+\sum_{n=1}^r
\{ {\cal T}_\alpha , \Pi_A \} \cdot  d_n  E^A_n  
-\!\sum_{\beta=1}^{4-r} \theta_{\alpha\beta} \,\cdot\, z_\beta 
\label{fifth}
\ee
which are $4-r$ relations linear in the remaining $4-r$ Lagrange multipliers $z_\alpha$. The DoF
thus depend on how many multipliers are determined in these conditions.  The matrix
$\theta_{\alpha \beta}$ is defined as
\be
{\cal \theta}_{\alpha \beta}
\equiv\;
\chi_{\alpha}^A(t,x) \,\{\S_A(t,x),\S_B(t,y)\}\,\chi_{\beta}^B(t,y) \, .
\ee
Notice that due to the non-trivial dependence on $x, \, y$,
$\theta_{\alpha \beta}$ is not necessarily antisymmetric and its rank
$s$ is not always even.
If $s<4-r$, some $z_\alpha$ are undetermined and we have $4-r-s$ new quartic constraints which
reduce the DoF. 

At the next (fifth) step, if the conservation of quartic constraints determines all the needed
multipliers, the procedure stops and we have only the $16-2r-s$ constraints $4\, (\Pi_A)+ 4 \,
(\S_A) +( 4-r)\, ({\cal T_\alpha})+(4-r-\,s) \, (\Q_\a)$. This implies that the maximal number of
states is
\be
 \D\le 4+r/2 \, .
\ee
This is the maximal number because if e.g.\ some of the $z_\a$ are not determined further steps are
necessary, possibly reducing the number of DoF.  In general also first class constraint may be
present corresponding to residual gauge invariances, further reducing the number of DoF.

\medskip

Summarizing, massive deformations with 5 DoF exist only in two cases: $r=3$, $s=0$ and $r=2$,
$s=2$. In the case $r=3$, the condition $s=0$ is equivalent to the differential equation (\ref{add})
for the deforming potential. In the case $r=2$ and $s=2$, (\ref{fifth}) determine the remaining two
LMs and all 10 constraints are consistent with the time evolution. Notice that it is not possible to
build a potential with $s=2$ without breaking spatial rotations.  Thus, we have shown that
$r=\text{Rank}(\V_{AB}) =3$, together with (\ref{add}) are necessary and sufficient conditions for a
rotational invariant massive gravity theory with 5 DoF (again we suppose that none of the found
constraints are trivial, i.e.\ $0 \approx0$).
 
\SEC{Solutions.} Let us then focus on how the $r=3$ condition can be realized. If spatial rotations
are preserved by the potential, the only option is
\be 
\det(\V_{AB})=0\,,\qquad \det(\V_{ij})\neq0  \, 
\label{master}
\ee
or equivalently ${\V}_{00}-\V_{0i}(\V_{ij})^{-1}\V_{j0}=0$.  Physically this means that once the
equation for the shifts $N^i$ are solved and inserted back in the potential, ${\cal V}$ will be
linear in $N$. When (\ref{master}) and (\ref{add}) are satisfied, $\V$ is a candidate for a theory
of massive gravity with 5 propagating DoF.  It is well known that perturbatively the number of
propagating DoF depends strongly of the background~\cite{diegolabega}. On the contrary, our analysis
is fully non perturbative, background independent and it is not sensitive on the way the deformation
is built.  For instance, it equally applies to the case where $\V$ is built from the invariants of
the matrix $g^{-1} f$, where $f$ is an arbitrary fixed metric~\cite{HGS} or is built out of set of
four scalar fields~\cite{Rubakov,dub}.


Let us look more closely to~(\ref{master}). As a general remark, it is a homogeneous Monge-Ampere
equations~\cite{MongeAmpere}; a particular solution is known in a closed form and, remarkably, also
the general solution can be given in parametric form.  Thus, there exist a large class of
potentials, taken as a function of lapse, shift and spatial metric $\V(N,N^i,\g_{ij})$ that
satisfies~(\ref{master}). Typically, one first finds a family of solutions of~(\ref{master}) and
then tries to enforce (\ref{add}).

In fact, it is not hard to find particular solutions of (\ref{master}), though we stress that most
of the solutions found will in general be LB.  For instance, as a simplification one can take $\V$
to be function of just two rotationally invariant combinations of lapse and shift, $x=N$,
$y=\g_{ij}N^i N^j$, i.e.\ $\V=\V(x,y)$.  In this case (\ref{master})\ reduces to
\be
2 \,y\, \left(\V_{xy}^2-\V_{yy}
   \V_{xx}\right)=\V_{y}\,\V_{xx} \, ,
\ee
for which an infinite class of solutions can be found by separation of variables.  An example is a
class of solutions that we will compare later with the LI ones
\be
\V=\b_1 \, \left[ (x+\b_2)^2 -  (y^{1/2}+\b_3)^2 \right]^{1/2}+\b_4\,x \, ,
\label{2ds}
\ee 
where $\b_{n=1,\ldots,4}$ are arbitrary scalar functions of $\g_{ij}$. Thus it is relatively simple
to construct solutions of (\ref{master}). Much more restrictive is condition (\ref{add}). In two
dimensions, the class of potentials (\ref{2ds}) satisfies also (\ref{add}) if $\beta_1= c_1$,
$\beta_2 = c_2 \gamma^{1/2}$, $\beta_4 = c_4 \gamma^{1/2}$ and $\beta_3$ is an arbitrary function of
the spatial metric $\gamma_{11}$, with $c_1$, $c_2$, $c_4$ constants.

\pagebreak[3]

More in general, it is known that in four dimensions the general solution of the Monge-Ampere
equation can be written implicitly in terms of two functions~\cite{MongeAmpere}.  When~(\ref{add})\
is also considered, one is left with infinitely many solutions in terms of a single function. Of
course most of the solutions will be Lorentz breaking.  An example is the following two-parameter
solution
\be
{\cal V}= m^2 \, N \, \gamma^{1/2} \left[c_1 \, g^{ij} \delta_{ij} + c_2  (g^{ij} \delta_{ij})^2 \right]\, ,
\ee
where $g^{ij}$ are the spatial components of the inverse 4d metric.  One can check that both
(\ref{master})\ and (\ref{add})\ are satisfied.

Generic LB solutions with five propagating DoF are attractive in their own respect.  Indeed,
expanding such a LB potential around flat space at quadratic level, one can easily get a theory
featuring a finite range (Yukawa-like) gravitational potential~\cite{Rubakov,dub}.  The breaking of
Lorentz invariance, taking place only in the gravitational sector, is not phenomenologically
excluded (and on the contrary is potentially observable in the forthcoming gravitational waves
experiments).  More interestingly, the resulting theory can be free from the vDVZ discontinuity and
remain perturbative at the solar system scale.  Note that on the contrary to avoid the vDVZ
discontinuity any LI theory needs to rely on the Vainshtein mechanism~\cite{Vainshtein}, which is
not yet completely understood. Even taking it for granted, the price to maintain LI is high: gravity
in the solar system is non-perturbative, making the theory hardly predictive.  Thus Lorentz breaking
seems to offer a healthier alternative. After all in the presence of sources Lorentz symmetry in the
gravitational sector plays little role if any. For instance, to check LI in the gravitational sector
one should measure the velocity of a graviton in (almost) flat space and compare it with the speed
of light.  We leave for a future work the complete study of the solutions of the above differential
equations leading to 5 DoF and of their phenomenology.

\SEC{LI solutions.} Let us now study in detail the problem of finding a ghost-free and Lorentz
invariant potential. The requirement of Lorentz invariance reduces considerably the possible
solutions of (\ref{master}). We require that the potential is built out of the invariants of
$X=g^{-1} \eta$, made out from the spacetime metric and a fixed Minkowski metric $\eta$, which
automatically guaranties that $\V$ is Lorentz invariant.  To start, let us consider the simplest
case of two dimensions.  The potential $\V$ is a function of the spatial metric $\g\equiv\g_{11}$,
the lapse $N$ and shift $N^1$, and eq.~(\ref{master})\ reads
\be
{\V}_{N^1 N^1}{\V}_{NN} - {\V}_{N N^1}^2  = 0 \, .
\label{mast2}
\ee
To single out the LI solutions, one observes that $\V=\V(X)$ is only function of the eigenvalues
$\lambda_{1,2}$ of the matrix $X$. Therefore, after expressing $N$, $N^1$ in terms of $\l_1$,
$\l_2$, $\g$,~(\ref{mast2}) must hold for any $\g$.  The resulting equation is cubic in it, so
that~(\ref{mast2}) splits into two branches of three differential equations 
\be
\begin{split}
&\mathcal{V}^{(2,0)}=-\frac{3}{2\,\lambda_1}\,\mathcal{V}^{(1,0)}\,,\quad
\mathcal{V}^{(0,2)}=-\frac{3}{2\,\lambda_2}\,\mathcal{V}^{(0,1)} \, ,\\[1ex]
&\qquad
\mathcal{V}^{(1,1)}=-\frac{\lambda_1^{3/2}\,\mathcal{V}^{(1,0)}\pm\lambda_2^{3/2}\,
\mathcal{V}^{(0,1)}}{2\,\lambda_1\,\lambda_2\,(\lambda_1^{1/2}\pm\lambda_2^{1/2})\,}.
\end{split}
\ee
Both can be solved in terms of simple functions:
\be
\V_{I,II}
= \frac{\alpha_1  \sqrt{\lambda_1\lambda_2}+\alpha_2 \, (\sqrt{\lambda_1} \pm\sqrt{\lambda_2} )
+\alpha_3}{ \sqrt{\lambda_1 \lambda_2}} ,
\label{2db1sol}
\ee 
with $\alpha_{1,2,3}$ are integration constants. These solutions also satisfy (\ref{add}).  Indeed,
they are special cases of~(\ref{2ds}), as can be seen by writing them as functions of $x$, $y$,
$\g$:
\bea \nonumber
\V_{I,II}&=&
\a_1+\a_2
\left(\left(\sqrt{\g}\pm x\right)^2-y\right)^{1/2}
+\a_3\,x\sqrt{\g} \,.
\label{mgen}
\eea
Being LI, they can also be written as functions of $X$:
\bea
\V_{I}&=&
\alpha_1 +\alpha_2\, \frac{\text{Tr}
 (X^{1/2})}{\sqrt{\det X }}+\frac {\alpha_3} {\sqrt{\det X }}\, ,
\label{2dofX1}
\\[1ex]
\V_{II}&=&\alpha_1 +\alpha_2\frac{\sqrt{\text{Tr}
 (X)-2 \sqrt{\det X }}}{\sqrt{\det X }}+\frac {\alpha_3} {\sqrt {\det X }}\, .\ 
\label{2dofX2}
\eea
The first is precisely the two-dimensional version of the ghost-free potential found
in~\cite{Gabadadze:2011}.  The solution corresponding to the $-$ sign on the other hand is a new one.
We stress that the two family of solutions do not overlap.  It can be shown that the theory with
$\V_{II}$ can not admit a Minkowski background, being non-analytic there. On the other hand $\V_I$
does.

The same strategy can be extended to $d=3$ and 4, though the computation becomes much more involved.
For $d=3$ from (\ref{master})\ we obtain a system of 9 partial differential equations which admit
four families of different solutions, generalizing~(\ref{2db1sol}) to the case of three eigenvalues,
with possible switching of signs.  In four dimensions there are eight solutions, which satisfy a
system of 125 partial differential equations.  The first solution coincides with the one found
in~\cite{Gabadadze:2011}, generalizing~(\ref{2dofX1}) to $d=4$. The other solutions are again
obtained by permuting the signs of the eigenvalue square roots, and generalize~(\ref{2dofX2}). Also
in the four dimensional case only $\V_{I}$ admits a Minkowski background.  This is not very
surprising, as the solution in~\cite{Gabadadze:2011} was obtained after a resummation of the
perturbative theory around Minkowski.  Due to the complicated coupled system of partially
differential equations, in $d=3$ and $d=4$ we cannot prove that the solutions found in the LI case
are unique, though the 2d analysis seems to suggest so.  In $d=3,4$ one can show that the
generalization of (\ref{mgen}) satisfy also (\ref{add}.  Comparing the LI with the LB cases it is
clear that Lorentz invariance is a severe constraint that cuts off most of the solutions of
(\ref{master}) even before dealing with (\ref{add}).

\pagebreak[3]

\SEC{Discussion and conclusions.}  We studied the Hamiltonian structure of a general massive
deformation of GR, which leads to non-perturbative and background-independent results.  In
particular, the analysis applies to both the LI and LB cases. The condition for having 5 DoF amounts
to a set of differential equations for the deforming potential: (\ref{master}), which is of
Monge-Ampere form, and (\ref{add}).  The picture that emerges is that Lorentz invariance
considerably restricts the potentials with 5 DoF: in the lower dimensional case $d=2$, a unique
solution admits a Minkowski background. In the $d=3,4$ we checked that the ghost free potential
proposed in~\cite{Gabadadze:2011} satisfies~(\ref{master}) and (\ref{add}). In general there are new
LI potentials, which however do not admit Minkowski as a background.  Phenomenologically, LI
theories are hard to handle due to their non-perturbative nature inside the Vainshtein radius, that
has to be larger than solar system to avoid the clash with the bending of light experiments.  As a
result, the theory is highly nonlinear and even predicting the motion of Mercury becomes
non-trivial.  On the other hand, in our investigation we have clearly shown that there are many
Lorentz breaking solutions of (\ref{master}) and (\ref{add}) which are candidates to be a consistent
and weakly coupled phase of massive gravity.  The analysis of the solutions of (\ref{master}) and
(\ref{add}) and their phenomenology will be given in a future work.

\end{document}